\documentclass[a4paper]{jpconf}

\usepackage{amsmath,amssymb}
\usepackage{subcaption}
\usepackage{cleveref}
\usepackage{cancel} 
\usepackage{graphicx}
\usepackage{amsfonts}
\usepackage{hyperref}

\DeclareMathOperator{\E}{\mathbb{E}}
\newcommand{\MET}{E_T^{\textrm{miss}}}
\newcommand{\METhat}{\hat{E}_T^{\textrm{miss}}}
\newcommand{\METvec}{\vec{E}_T^{\textrm{miss}}}

\usepackage[compact]{titlesec}
\titlelabel{\thetitle.\hskip0.5em\relax}
\titlespacing{\section}{0pt}{0.5\baselineskip}{0.05\baselineskip}
\titlespacing{\subsection}{0pt}{0.5\baselineskip}{0.05\baselineskip}

\begin{document} 

\title{Particle Generative Adversarial Networks for full-event simulation at the LHC and their application to pileup description}

\author{Jes\'us Arjona Mart\'inez$^{1,2,3}$, Thong Q Nguyen$^2$, Maurizio Pierini$^3$, Maria Spiropulu$^2$ and Jean-Roch Vlimant$^2$}
\address{$^1$ University of Cambridge, Trinity Ln, Cambridge CB2 1TN, UK}
\address{$^2$ California Institute of Technology, 1200 E.\ California Blvd, Pasadena, CA 91125}
\address{$^3$ CERN, CH-1211 Geneva, Switzerland}
\ead{ja618@cam.ac.uk, \{thong, smaria, jvlimant\}@caltech.edu, maurizio.pierini@cern.ch}

\begin{abstract}
We investigate how a Generative Adversarial Network could be used to generate a list of particle four-momenta from LHC proton collisions, allowing one to define a generative model that could abstract from the irregularities of typical detector geometries. As an example of application, we show how such an architecture could be used as a generator of LHC parasitic collisions (pileup). We present two approaches to generate the events: unconditional generator and generator conditioned on missing transverse energy. We assess generation performances in a realistic LHC data-analysis environment, with a pileup mitigation algorithm applied.
\end{abstract}

\section{Introduction}

The simulation of subatomic particle collisions, their subsequent detector interaction, and their reconstruction is a computationally demanding task for the computing infrastructure of the experiments operating at the CERN Large Hadron Collider (LHC). The high accuracy of state-of-the-art Monte Carlo (MC) simulation software, typically based on the \textsc{GEANT4}~\cite{GEANT4}, has a high cost: MC simulation amounts to about one half of the experiments' computing budget and to a large fraction of the available storage resources~\cite{computingresources}, the other half being largely used to process simulated and real data (event reconstruction).


Following their invention in 2014, GANs gained traction as generative models, often superior to Variational Autoencoders~\cite{Variationalautoencoder} and with very impressive results in image production~\cite{CelebrityAINvidia, TransferlearningpaintingcycleGAN}. 
Due to their high inference speed, GANs can be used as fast-simulation libraries. This approach has been successfully investigated with proof-of-principle studies related to particle showers in multilayer calorimeters~\cite{CaloGANdeoliveira,NIPSLCD,Erdmann:2018jxd} and particle jets~\cite{Musella:2018rdi}, as well as in similar application to different HEP domains~\cite{Erdmann:2018kuh, LAGAN, LHCbGAN, QCDGAN, dijetGAN, anaspecGAN}. All these studies formalized the simulation task in terms of either image generation or analysis-specific high-level features. 

In this work, we present pGAN, a full-event particle-based generative model that can be used to emulate pileup simulation at the LHC.

\section{Pileup simulation}
\label{sec:usecase}

The majority of LHC proton-proton collisions result in so-called {\it minimum-bias} (MB) events, i.e., in low-energy ({\it soft}) interactions between proton constituents. These events are characterized by low-$p_T$ particles, as opposed to the head-on collision processes typically studied at the LHC (so-called {\it hard} or high-$p_T$ interactions).
Any hard interaction happens simultaneously to many parasitic MB events, generically referred to as {\it pileup}. Pileup simulation is a fundamental aspect of a realistic LHC simulation software. The current implementation of pileup simulation consists in overlapping a set of MB events to the main high-$p_T$ collision. Events could be generated on demand or be sampled from a pre-generated library. The former is computationally expensive, while the latter is inflexible and suffers from I/O issue.

GAN emerges as a possible solution to speed up the on-demand generation of MB events and remove the need for a pre-generated library. To our knowledge, the only application of machine learning to pileup simulation is the work presented in Ref.~\cite{PileupDCGAN}, where pileup images are generated using a Deep Convolutional GAN model (DCGAN)~\cite{DCGAN}. However, an image-based event representation can't be used as input for downstream reconstruction algorithms. In addition, our proposed pGAN, which uses particle-based event representation, can abstract from the details of the detector geometry (e.g., its irregularities) and better scales with the foreseen increase of detector complexity. 

\section{Dataset}
\label{sec:dataset}

Synthetic MB events from proton-proton collisions are produced using the {\tt PYTHIA8}~\cite{Pythia} event generator. The center-of-mass energy of the collision is set to 13~TeV, corresponding to the LHC Run II (2015-2018).
All soft QCD processes are activated, allowing for both initial- and final-state radiation as well as multiple parton interactions. The produced events are passed to {\tt DELPHES}~\cite{Delphes}, to emulate the detector response. We take as a reference the upgraded design of the CMS detector, foreseen for the High-Luminosity LHC phase. 
The {\tt DELPHES} particle-flow reconstruction algorithm is applied, returning the list of charged particles, photons, and neutral hadrons in the event. 
Minimum bias events are then combined to simulate a per-event pileup contribution, with mean number of MB events $\overline{n}_{\textrm{PU}} = 20$ following a Poisson distribution. We randomly sample $n_{\textrm{PU}}$ events from the MB dataset and mix them by merging the list of charged particles, neutral hadrons, and photons across the events.

Due to the complexity of training on long sequences, we restrict a maximum of 150 particles per event: the 50 charged particles, 50 photons and 50 neutral hadrons with the highest $p_T$ value. This choice is mainly due to technical limitations that a more powerful training setup might help to overcome. On the other hand, cutting the sequence after $p_T$ ordering is a well motivated simplification of the problem: a typical physics analysis would be based on a pileup mitigation algorithm, which usually removes the majority of the soft pileup contamination. 




\section{Network architectures}
\label{sec:network}

A GAN consists of two neural networks, a generator $\mathcal{G}$ and a discriminator $\mathcal{D}$. Given a set of samples $x$, the aim of a GAN training is to learn the function $p_{\textrm{data}}(x) \in A$ under which the $x$ samples are distributed. We define an $n$-dimensional prior of input noise $z \sim p_z(z) \in \mathbb{R}^n$. The generator $\mathcal{G}$ is a differentiable function with trainable parameters $\theta_{\mathcal{G}}$, mapping $\mathbb{R}^n$ to $A$. The discriminator $\mathcal{D}$, with trainable parameters $\theta_{\mathcal{D}}$, is a map between $A$ and $[0, 1]$, returning the probability that a given sample belongs to the set of real samples rather than originating from $\mathcal{G}$. $\mathcal{D}$ is trained to assign the correct probability to both real and generated ("fake") data; $\mathcal{G}$ is trained to produce samples such that they maximize the probability of them being real $\mathcal{D}(\mathcal{G}(z))$. 

We develop 2 models for pGAN: an unconditional pGAN where the generator starts from purely random noise $z$, and a conditional pGAN ~\cite{CGAN} where a label acts as an extension of $z$ to allow for generation of events based on an initial condition. In our use case, the missing transverse energy $\MET$ is chosen as the label due to its importance in most physics analyses.

The loss function to train $\mathcal{G}$ and $\mathcal{D}$ in an unconditional pGAN is described as:
\vspace*{-0.3\baselineskip}
\begin{equation}
\mathcal{L}^{\textrm{uncond}} = \mathcal{L}_{\textrm{adv}} = \E_{z\sim p_z(z)}[\log(\mathbb{P}(D(G(z))=0)] + \E_{I\sim A}[\log(\mathbb{P}(D(I)=1))]~,
\vspace*{-0.2\baselineskip}
\end{equation}
while for conditional pGAN, the loss function becomes:
\vspace*{-0.3\baselineskip}
\begin{gather}
\mathcal{L}^{\textrm{cond}} = \mathcal{L}_{adv} + \alpha \mathcal{L}_{aux}~, \textrm{~where} \\
\mathcal{L}_{\textrm{aux}} = \E_{(z, \MET) \sim (p(z) \times f(A))} 
\left[\frac{|\MET - \METhat(G(z|\MET))|}{\MET}\right] + \E_{x\sim A}\left[\frac{|\MET(x)-\METhat(x)|}{\MET(x)}\right]
\end{gather}
$\METhat$ is the missing transverse energy value computed from list of particles input to the discriminator, $f(A)$ is the empirical $\MET$ distribution of the dataset. In the conditional GAN setting, the dataset is transformed such that $\phi(\METvec)=0$ and the $\phi$ value of each particle is computed as azimuth angle between its momenta and $\METvec$. This way a single scalar value $\MET$ can describe the full vector $\METvec$ since the direction is chosen as the coordinate axis.





\begin{figure}[t]
\centering 
\begin{subfigure}{0.49\textwidth}
\centering
\includegraphics[width=0.99\textwidth]{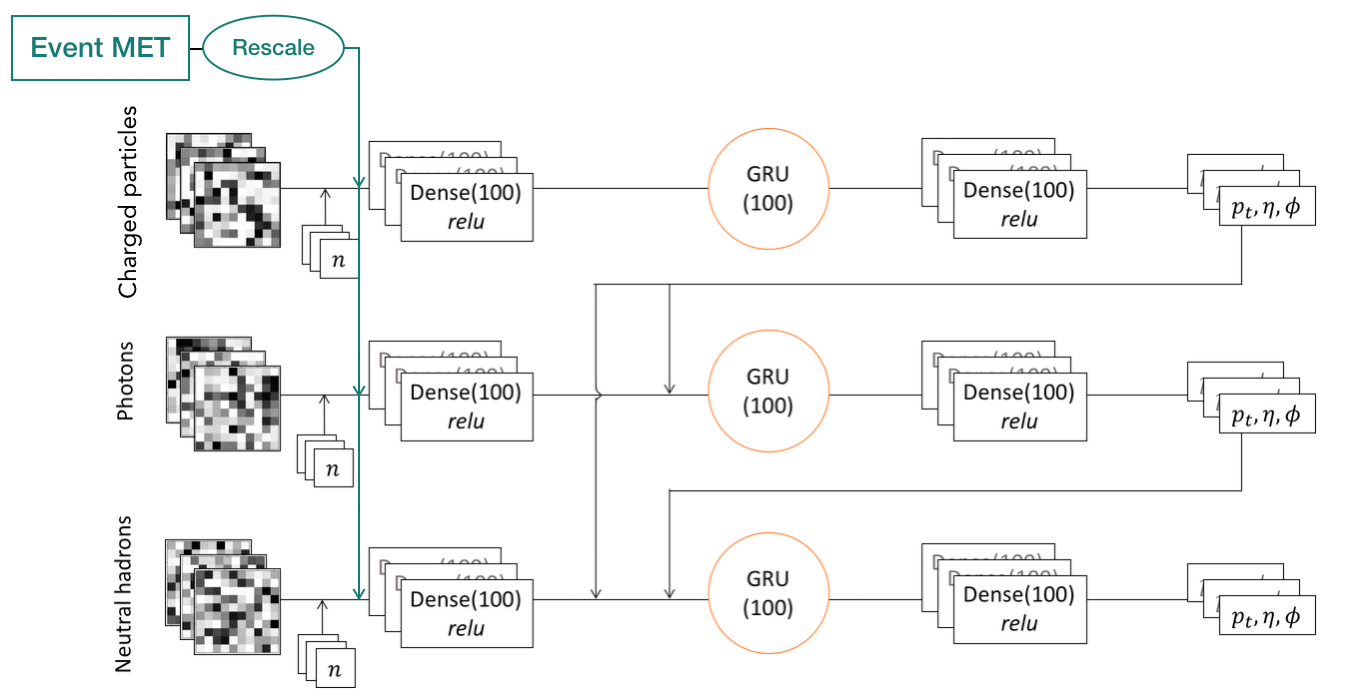}

\end{subfigure}
\hfill
\begin{subfigure}{0.49\textwidth}
\centering
\includegraphics[width=0.99\textwidth]{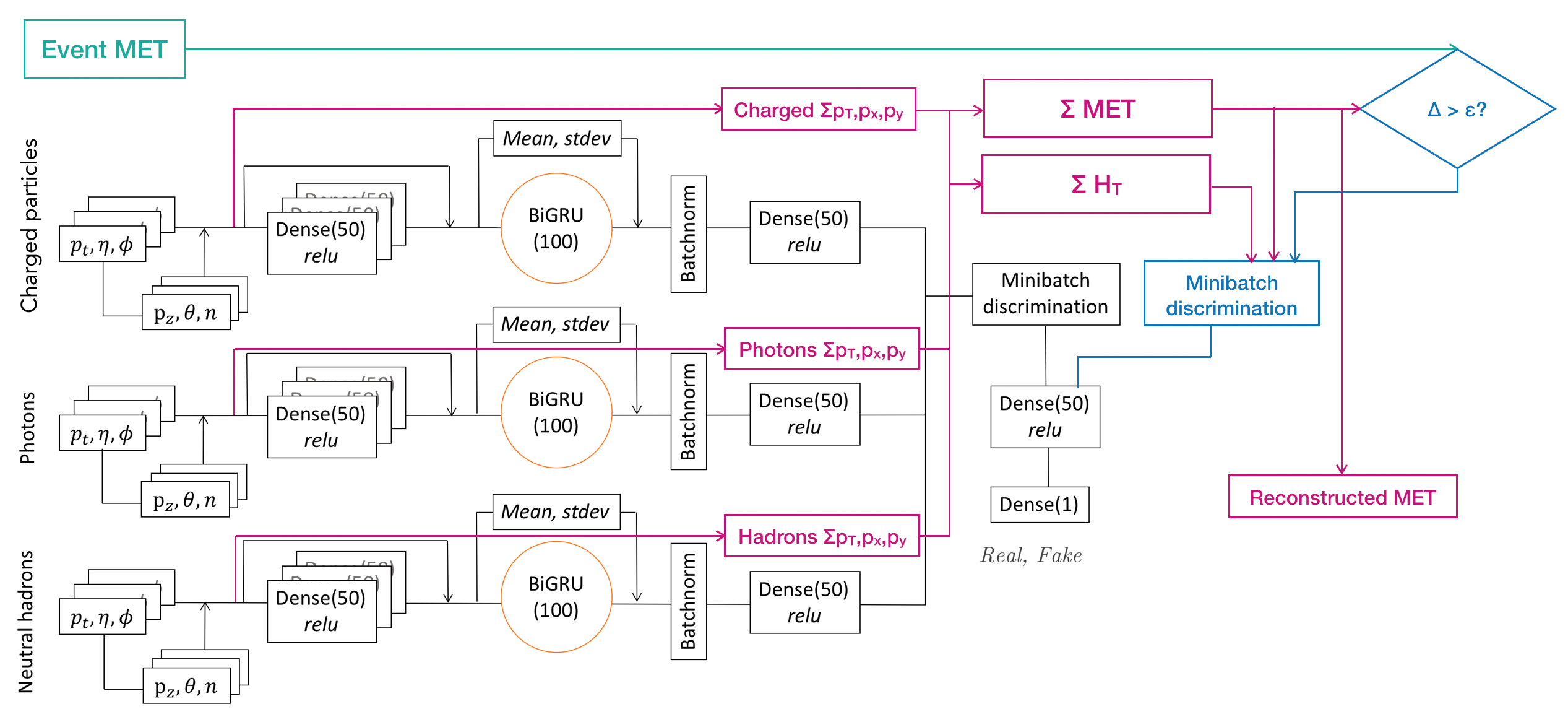}
\end{subfigure}
\caption{The architecture of conditional pGAN: generator $\mathcal{G}^{\textrm{cond}}$ (left) and discriminator $\mathcal{D}^{\textrm{cond}}$ (right). Arrows signify concatenation. Details are described in the text.}
\label{fig:architecture} 
\end{figure}

\subsection{Unconditional pGAN}
\label{sec:UpGAN}

\subsubsection{Generator}

The $\mathcal{G}$ model takes as input a set of time steps, corresponding to the number of particles to generate, sampled from a uniform distribution from 0 to 50 for each particle class. Along with the noise, the current particle number $n$ is given to the network.  We represent each particle as a $(p_T, \phi, \eta)$ tuple. 
We enforce the $\phi$-symmetry constraint through a mod-$2\pi$ activation function for $\phi$. The input in each branch from the noise input is processed via a block of Dense layers, a GRU layer, and another block of Dense layers as the final outputs. Each block of Dense consists 3 concatenated Dense layers that represent the particle's $p_T$, $\phi$, and $\eta$ respectively. 


Due to the peculiar shapes of the $\eta$ distributions, we pre-train a small dense network that fed with a Gaussian-distributed input mimics the $\eta$ distribution as output. This network is used to process the $\eta$ output of the generator. This can be viewed as an activation function parametrizing the $\eta$ distribution in an unbiased way and decouples learning the distribution from the already complex particle generation task. 

 \subsubsection{Discriminator}
 The discriminator  $\mathcal{D}$ is of a binary classifier whose first step consists in a Physics layer, which takes each particle's defining features ($p_T,\phi, \eta, n$) and concatenates to them other redundant features ($\theta, p_z, p$). This Physics layer is introduced to maximize the information given to the discriminator without adding redundant information to the particle representation returned by the generator. This prevents the generator for having to learn dependencies between different features of the particle representation, while allowing the discriminator to exploit them. 
 The discriminator makes use of a Bidirectional recurrent layer, which reduces the dependence on the long term memory of the GRU cell. A layer that calculates the mean and standard deviation of the features in a dense layer is included, in a similar spirit to feature matching~\cite{ImprovedtechniquesforGANs}. Mini-batch discrimination is used to prevent mode collapse.
 

\subsection{Conditional pGAN}
\subsubsection{Generator}
The generator $\mathcal{G}^{\mathrm{cond}}$ for the conditional pGAN, as shown in Fig.~\ref{fig:architecture}, is built on top of the unconditional generator $\mathcal{G}$. An initial value of the $\MET$, sampled from real data, is injected as the input of the generator along with the noise after being rescaled to same range of the noise. This initial $\MET$ is also used as the input to the discriminator. The rest of the generator architecture is similar to the unconditional pGAN's $\mathcal{G}$.

\subsubsection{Discriminator}
The discriminator $\mathcal{D}^{\mathrm{cond}}$, as shown in Fig.~\ref{fig:architecture}, takes as inputs the 3 lists of particles along with the event $\MET$, which is either the initial condition $\MET$ (in the case of the generated lists of particles) or the actual $\MET$ (in the case of sampling the lists of particles from the training data). In addition to usual computation flow in the unconditional $\mathcal{D}$ described in Sec.~\ref{sec:UpGAN}, $\mathcal{D}^{\mathrm{cond}}$ also computes a few high-level features out of the input particle lists, in particular the reconstructed $\MET$ and $H_T$ (the scalar sum of all input particles' momenta), which are used as inputs to the final prediction. A binary flag $\Delta$, which returns 1 if the absolute difference between reconstructed $\MET$ and the initial $\MET$ is greater than $\epsilon$ and returns 0 otherwise, is also concatenated to the input of the final prediction. This can be viewed as an attempt to let the discriminator to learn some global kinematic features of the inputs. Additionally, the reconstructed $\METhat$ are compared with the initial input $\MET$ to construct the additional term $\mathcal{L}_{aux}$ in the loss function, as described in Eq.~(3). 

\subsection{Implementation details}
Adam~\cite{Adam} is used for optimization with a batch size of 32, a learning rate of $1 \times 10^{-4}$ for the generator and of $2 \times 10^{-4}$ for the discriminator. For the conditional pGAN, we choose $\epsilon=10$ GeV and $\alpha = 0.05$. We train on cropped sequences of variable length, forcing the discriminator to learn to distinguish even very short arrays, and make use of dropout in the discriminator. Batch normalization~\cite{Batchnorm} is included in the discriminator but not in the generator. Hyperparameter optimization was performed on the learning rate of both networks. We find Least-Squarse GAN~\cite{LeastSquaresGANLSGAN} offers the most stable training, outperforming both Wasserstein GAN~\cite{WassersteinGAN} and the original GAN implementation.

\section{Experimental evaluation}
\label{sec:evaluation}

GAN performance evaluation is a challenging task. Due to their unsupervised nature, domain and task-specific metrics are often required. Our evaluation technique is three-fold: (i) we plot distributions of relevant physical features and quantify the matching between the ground-truth (GT), {\tt PYTHIA8} + {\tt DELPHES}, and our network; (ii) we make use of high-level global event-shape variables such as the transverse thrust and the $\MET$; (iii) we evaluate the effect of using our proposed generation technique in a real analysis environment. To this purpose, we apply a state-of-the-art pileup removal algorithm ({\it SoftKiller}~\cite{Softkiller}) and cluster the remaining particles in the event, characterizing the agreement between real simulation and pGAN on jet kinematic properties (e.g., the jet $p_T$). These three different performance assessments are discussed in the rest of this section.

  
\subsection{Histogram matching} 

\begin{figure}[t]
\centering 
\begin{subfigure}{0.9\textwidth}
\includegraphics[width=.32\textwidth]{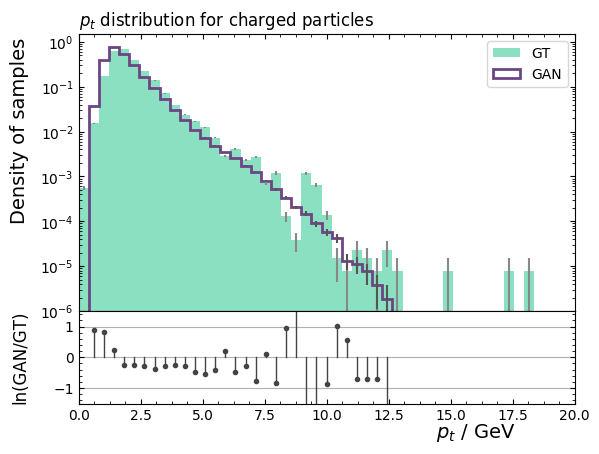}
\includegraphics[width=.32\textwidth]{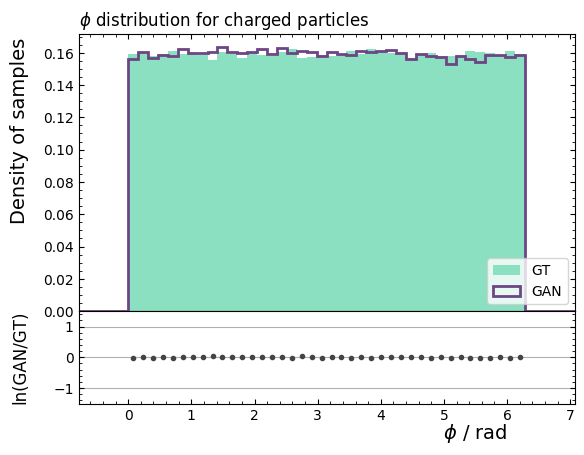}
\includegraphics[width=.32\textwidth]{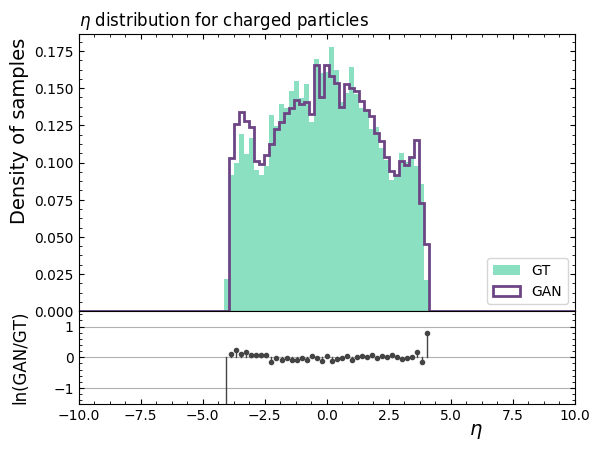}
\end{subfigure}

\centering 
\begin{subfigure}{0.9\textwidth}
\includegraphics[width=.32\textwidth]{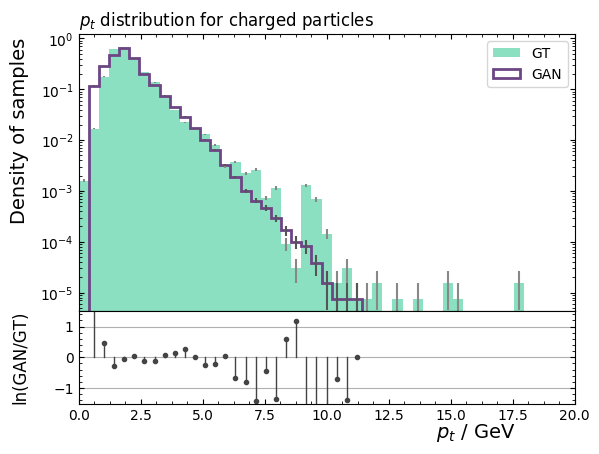}
\includegraphics[width=.32\textwidth]{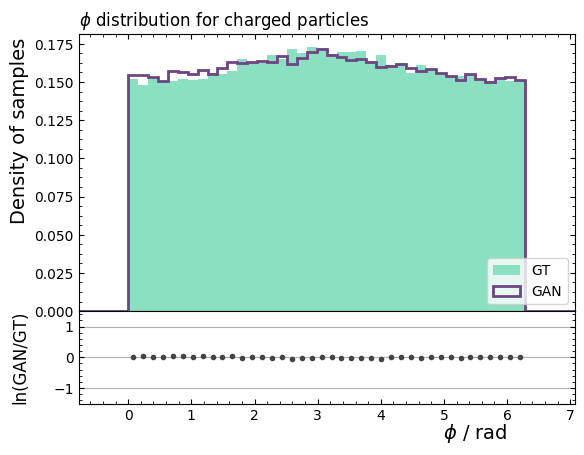}
\includegraphics[width=.32\textwidth]{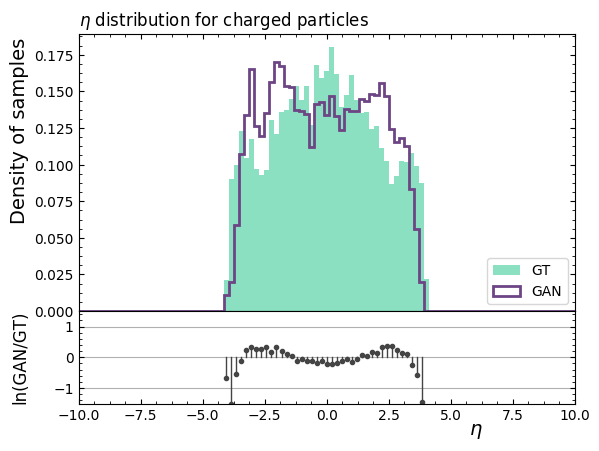}
\end{subfigure}

\vspace*{-0.6\baselineskip}
\caption{\label{fig:histograms} Comparison of the transverse momentum $p_T$ (left), azimuth angle $\phi$ (center) and pseudorapidity $\eta$ (right) for charged particles between the test data and the events generated by unconditional pGAN (top) and conditional pGAN (bottom). For the conditional pGAN, $\phi$ is transformed to be the azimuth angle between the particles' momenta and $\METvec$.} 
\end{figure}


   
\begin{figure}[thpb]
\centering 
\begin{subfigure}{0.48\textwidth}
\includegraphics[width=.48\textwidth]{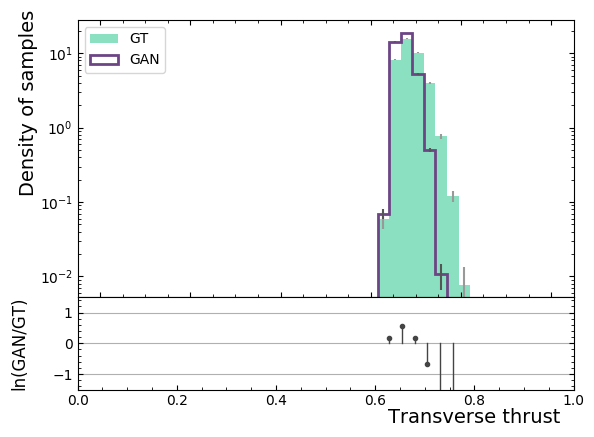}
\includegraphics[width=.48\textwidth]{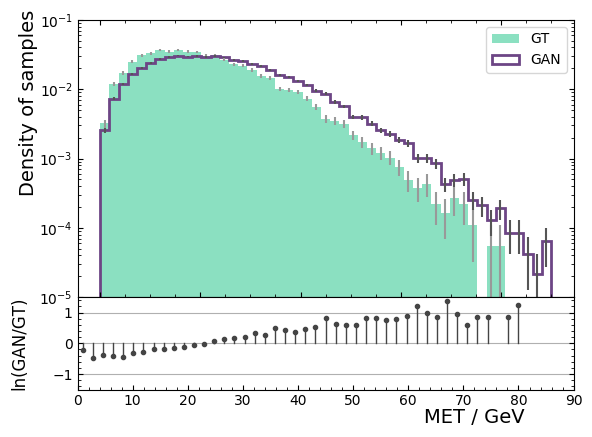}
\vspace*{-0.4\baselineskip}
\caption{Unconditional pGAN}
\end{subfigure}
\begin{subfigure}{0.48\textwidth}
\includegraphics[width=.48\textwidth]{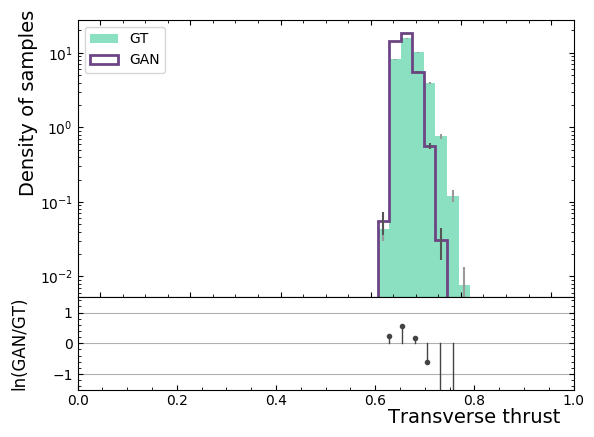}
\includegraphics[width=.48\textwidth]{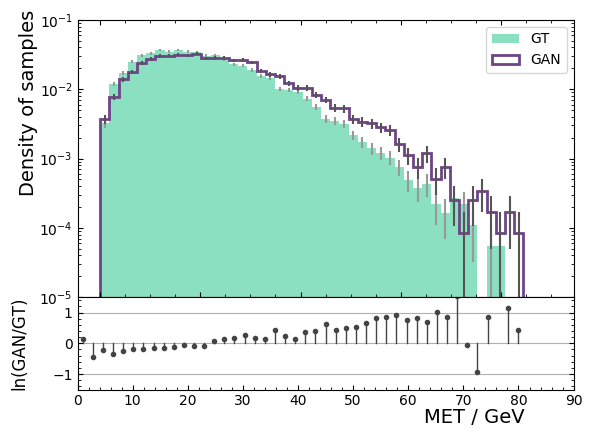}
\vspace*{-0.4\baselineskip}
\caption{Conditional pGAN}
\end{subfigure}

\vspace*{-0.5\baselineskip}
\caption{\label{fig:highlevel} Comparison of the transverse thrust and $\MET$ distributions between the test data and the generated events by pGANs.} 
\end{figure}

Fig.~\ref{fig:histograms} shows how the pGAN generators succeed in learning the main aspects of the particles' $p_T$, $\eta$ and $\phi$ distributions. The comparison is limited to the first 50 high-$p_T$ particles of each class, in order for the representation of the generated event to be consistent with the GAN generator output. 
We observe a remarkable agreement in $p_T$ and $\phi$: the long tail of the transverse momentum distribution is well described across four orders of magnitude and the $\phi$ rotation invariance of the physical process is reproduced. On the other hand, while the qualitative features of the pseudorapidity distribution are learned, the agreement is not completely satisfactory. A better match should be the goal of future work. 
Fig.~\ref{fig:highlevel} shows comparisons for some of the event-related high level features used in model evaluation: transverse thrust and $\MET$. We observe some discrepancy being associated to overall scale shifts, related to the different truncation criteria applied to the two sets of events. 

\begin{figure}[thpb]
\centering 
\includegraphics[width=0.9\textwidth]{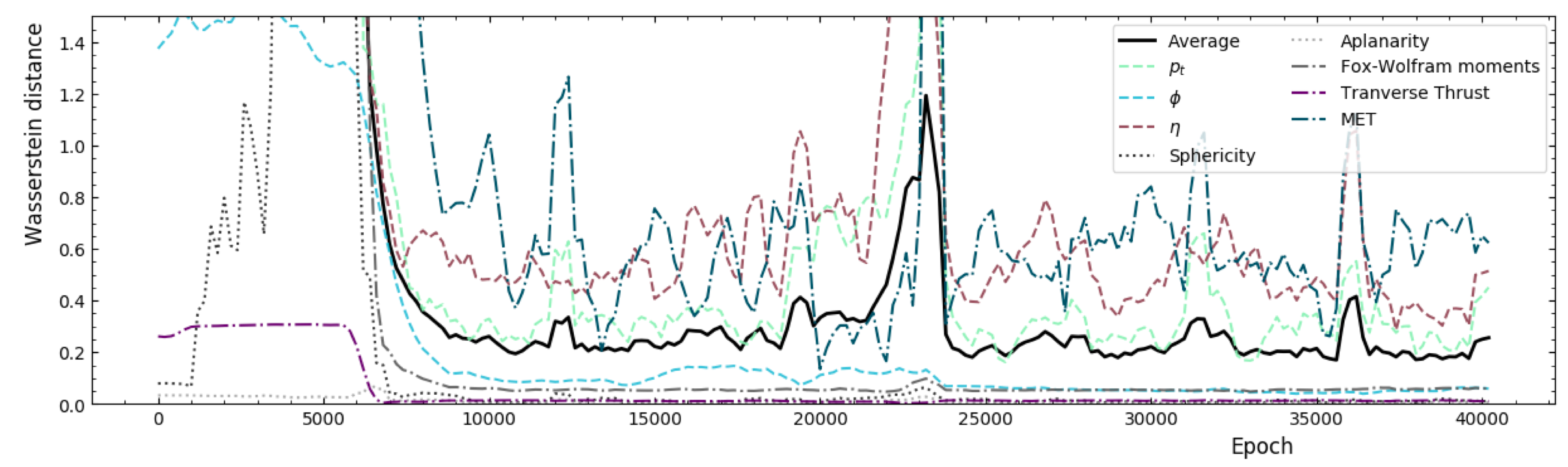}
\hfill
\vspace*{-0.6\baselineskip}
\caption{\label{fig:loss} Evolution of our performance  metric (solid black) as a function of training. EM distances for some of the individual quantities are superposed.}
\end{figure}

\subsection{Wasserstein metric}

We use the previously described histograms to quantify the difference between the target and generated distributions through the Wasserstein or Earth Mover's (EM) distance. 
We define our performance metric as the weighted average of the EM distances over the feature distributions: $p_T$, $\eta$, $\phi$ for all three particle types, sphericity, transverse thrust, aplanarity, $\MET$, and the first and second transverse Fox-Wolfram moments~\cite{Fox-Wolframmoments}. The use of this metric tackles the problem of lack of interpretability of the loss function: we observe that the metric decreases steadily as the training progresses, as shown in Fig.~\ref{fig:loss}, providing a way of monitoring progress, performing early-stopping and tracking training failure. Based on this metric, we perform model comparison, hyperparameter tuning, and the final best-model choice.

\subsection{Pileup subtraction}
Typical LHC analyses are performed after applying a pileup removal algorithm, which aims to subtract soft radiation from QCD. It is then important to demonstrate that the pGAN is good in modeling the residual pileup contribution, after such a subtraction algorithm is applied. Since this residual pileup contribution is the only relevant effect for physics analyses, it is acceptable for a pileup emulation software to have a non-accurate pileup simulation as long as the pileup effect is well model after applying a pileup mitigation technique. For this purpose, we consider a sample of $Z \rightarrow \nu \overline{\nu}$ events, generated using {\tt Pythia8}. Events are processed with {\tt Delphes}, using the same setup adopted to generate the pileup reference sample, both with and without activating the pileup emulation  at $\overline{n}_{\textrm{PU}} = 20$. The generated no-pileup events are mixed with the pileup emulation returned by the generator. The {\it SoftKiller}~\cite{Softkiller} algorithm with a grid size of $a \approx 0.5$ is then applied to  both these events and those with a full pileup simulation. 



\begin{table}[h]
\caption{\label{tab:1} Mean leading-jet $p_T$ for events with no pileup and pileup generated by {\tt Pythia8} and by the network (pGAN), before and after running {\it SoftKiller}.}
\vspace*{-0.6\baselineskip}
\begin{center}
\begin{tabular}{rcrc}
\br
                             & $\langle p_T \rangle$ / GeV & & $\langle p_T \rangle$ / GeV\\
\mr
No PU                        & 136.8  & &           \\
Pileup - GT              & 146.6  & Pileup - GT - subtracted    &  135.0          \\
Pileup - pGAN                 & 141.1 &  Pileup - pGAN - subtracted   &  135.7             \\
\br
\end{tabular}
\end{center}
\end{table}

\begin{figure}[h]
\centering 
\begin{subfigure}{.4\textwidth}
\includegraphics[width=.99\textwidth]{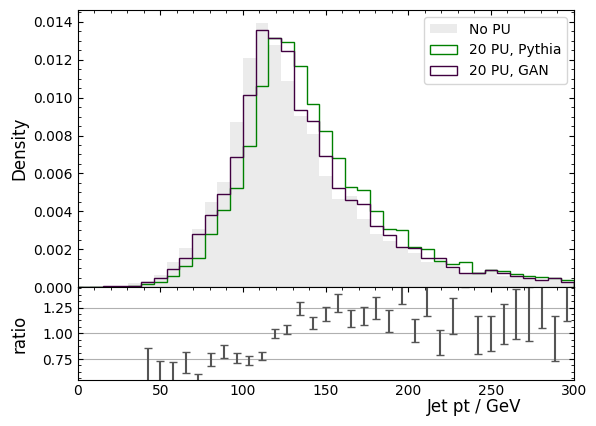}
\end{subfigure}
\begin{subfigure}{.4\textwidth}
\includegraphics[width=.99\textwidth]{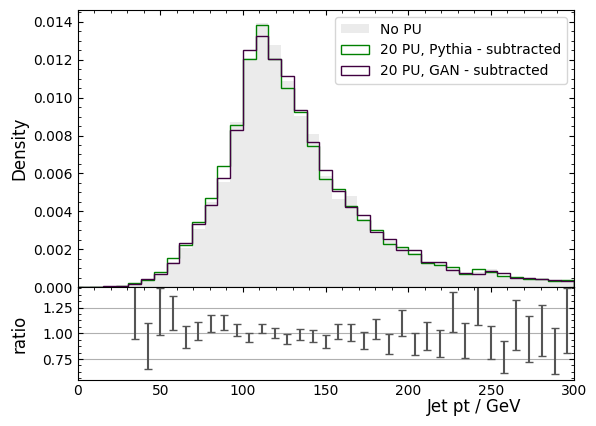}
\end{subfigure}
\vspace*{-0.6\baselineskip}
\caption{\label{fig:softkiller} Comparison between leading jet $p_T$ distributions for events with no pileup (solid black) and pileup generated by {\tt Pythia8} (green) and by the network (magenta). Distributions are shown both before (left) and after (right) running the {\it SoftKiller} pileup mitigation algorithm. The bottom plot shows the Pythia/GAN ratio.}
\end{figure}

Fig.~\ref{fig:softkiller} shows the $p_T$ distribution of the highest-$p_T$ jet in the event in various configurations. The main effect of pileup contamination is a shift in the $p_T$ distribution towards larger values. The shift is underestimated when the pileup is described through the pGAN generator, which is due to the fact that pGAN only returns the first $150$ particles per event, instead of the usual $\sim 900$. After processing the event with the {\it SoftKiller} algorithm, the leading jet $p_T$ distribution for {\tt Pythia8} and our GAN match within $0.7~\textrm{GeV}$. The agreement could be further improved by increasing the number of neutral hadrons and charged particles returned by the pGAN. 

\section{Conclusions}
\label{sec:conclusions}

We presented GAN models based on a recurrent unit, capable of generating lists of particles with variable lengths and also with an initial condition on $\MET$. Such a model could be used for particle-based simulation software, such as those of experiments using particle-flow reconstruction algorithms. This model could be used to replace ordinary rule-based algorithms in specific aspects of jet generation. In this paper, we show its application to pileup emulation in LHC collisions. While technical limitations forced us to reduce the length of the returned particle chain, the network is capable of emulating the effect of pileup on a realistic data analysis, after applying a pileup mitigation algorithm. 
 

\section*{Acknowledgments}
We are grateful to  Caltech and the Kavli Foundation for their support of undergraduate student research in cross-cutting areas of machine learning and domain sciences. This work was conducted at  "\textit{iBanks}", the AI GPU cluster at Caltech. We acknowledge NVIDIA, SuperMicro  and the Kavli Foundation for their support of "\textit{iBanks}".
This project has received funding from the European Research Council (ERC) under the European Union's Horizon 2020 research and innovation program (grant agreement n$^o$ 772369). This project is partially supported by the United States Department of Energy, Office of High Energy Physics Research under Caltech Contract No. DE-SC0011925.

\section*{References}
\bibliographystyle{iopart-num}
\bibliography{iopart-num}

 

\end{document}